\documentstyle[prl,aps,preprint]{revtex}

\def\mxth{\mathsurround=0pt }
\def\xversim#1#2{\lower2.pt\vbox{\baselineskip0pt \lineskip-.2pt
\ialign{$\mxth#1\hfil##\hfil$\crcr#2\crcr\sim\crcr}}}

\def\lta{\mathrel{\mathpalette\xversim <}}
\def\gta{\mathrel{\mathpalette\xversim >}}

\def\hmpc{{\rm\, h^{-1}Mpc}}

\def\kms{{\rm\, km\ s^{-1}}}

\def\mpc{{\rm\,Mpc}}

\def\'{^{\prime}}
\def\avrg#1{{\langle #1 \rangle}}

\def\Dt{\spose{\raise 1.5ex\hbox{\hskip3pt$\mathchar"201$}}}    
\def\dt{\spose{\raise 1.0ex\hbox{\hskip2pt$\mathchar"201$}}}    
\def\spose#1{\hbox to 0pt{#1\hss}}

\def\lta{\mathrel{\spose{\lower 3pt\hbox{$\mathchar"218$}}
     \raise 2.0pt\hbox{$\mathchar"13C$}}}
     \def\gta{\mathrel{\spose{\lower 3pt\hbox{$\mathchar"218$}}
	  \raise 2.0pt\hbox{$\mathchar"13E$}}}

	  \def\eg{{\it e.g., }}
	  
	  \def\etal{{\it et al. }}

\begin{document}
\draft
\title{The Imprint of Gravitational Waves
on the Cosmic Microwave	Background}
\author{Robert Crittenden,$^1$
J. Richard Bond,$^2$
Richard L. Davis,$^1$
George Efstathiou,$^3$ Paul J. Steinhardt$^1$}
\address{$^{(1)}$  Department of Physics, University of Pennsylvania,
Philadelphia, PA  19104 \\
$^{(2)}$  Canadian Institute for Theoretical Astrophysics,
University of Toronto,
Toronto, Ontario, Canada
M5S 1A7  \\
$^{(3)}$  Department of Astrophysics,
 Oxford University, Oxford, England  OX1 3RH }
\maketitle
\begin{abstract}
Long-wavelength gravitational waves can induce significant temperature
anisotropy in the cosmic microwave background.  Distinguishing this
from anisotropy induced by energy density fluctuations is critical for
testing inflationary cosmology and theories of large-scale structure
formation.  We describe full radiative transport calculations of the
two contributions and show that they differ dramatically at angular
scales below a few degrees. We show how anisotropy experiments probing
large- and small-angular scales can combine to distinguish the imprint
due to gravitational waves.
\end{abstract}

The cosmic microwave background (CMB) temperature anisotropy may be
induced by energy density fluctuations {\it and} by gravitational
waves\cite{AWSta}, corresponding to scalar and tensor metric
perturbations, respectively.  Although anisotropy measurements probing
angular scales above a few degrees (\eg COBE\cite{dmr}) cannot
discriminate scalar from tensor\cite{Dav}, we show in this Letter that
the two contributions can be separated when data from smaller angle
experiments are used as well.

Resolving the two contributions relies upon detailed theoretical
predictions for the form of the multipole components, $a_{\ell
m}^{(S)}$ and $a_{\ell m}^{(T)}$, of the relative temperature pattern
on the sky, $\Delta T/T (\theta , \phi)$.  For inflationary models,
each multipole for the two modes is predicted to be statistically
independent and Gaussian-distributed, fully specified by angular power
spectra, $C_\ell^{(S)}=\avrg{\vert a_{\ell m}^{(S)}\vert^2 }$ and
$C_\ell^{(T)}=\avrg{\vert a_{\ell m}^{(T)}\vert^2 }$. Although
$C_\ell^{(S)}$ has been calculated before\cite{BE}, $C_\ell^{(T)}$ was
previously known only for low multipoles, $\ell \lta 30$, relevant for
angles above a few degrees\cite{AWSta,Dav,Oth}. For these $\ell$'s,
the dominant spatial wavelengths contributing to $C_\ell^{(T)}$ and
$C_\ell^{(S)}$ were outside the horizon at photon decoupling, and both
scalar and tensor modes induce similar red shifts and blue shifts in
the CMB\cite{AWSta,Dav,Oth}. For example, COBE's DMR is unable to
distinguish the two contributions due to cosmic variance (from the
theory signal) and experimental noise, although their sum can be
determined from one year of DMR data to within 30\% (see Fig.~2a
below), improving to about 15\% with four full years of data.

In this Letter, we compute $C_\ell^{(T)}$ to much higher multipoles,
and show that the predicted $C_\ell^{(T)}$ becomes highly suppressed
relative to $C_\ell^{(S)}$ at large $\ell$.  The dominant wavelengths
for $\ell \gta 30$ were inside the horizon at decoupling.  Inside the
horizon, scalar-mode anisotropies are enhanced by the gravitational
instability of density perturbations and by Thomson scattering from
moving electrons, whereas gravitational waves disperse as freely
propagating, massless excitations and red shift away.  Taking
advantage of this difference, we find that combining experiments at
small and large angular scales can determine the scalar and tensor
components.  Current CMB anisotropy data at small scales is not yet
good enough to do so, especially since some of the signals observed
may be Galactic rather than cosmic in origin, but the statistical
errors of these and other near-future experiments are small enough to
allow separation at about the two sigma level, as we show here.

Separating tensor from scalar is essential for theories of cosmic
structure formation since it is from $C^{(S)}_\ell$ that we can infer
the amplitude of the primordial density fluctuations. It also provides
a critical test for inflationary cosmology\cite{Dav,Oth,Abffo}. All
inflation models produce a post-inflation spectrum of scalar and
tensor metric fluctuations with some tilt away from the
scale-invariant (spectral index $n_s=1$) form.  The degree of tilt
depends upon the details of the inflation model but may be quite
significant\cite{Dav,Oth,Abffo} ($n_s \gta 0.5$). For a wide range of
inflation models, but not all, the gravitational wave content of the
CMB anisotropy is related to the tilt\cite{Dav,Oth,Abffo}:
\begin{equation}
C^{(T)}_2/ C^{(S)}_2 \approx 7(1-n_s) \ ,
\end{equation}
where $C^{(T)}_2$ and $C^{(S)}_2$ are the $\ell =2$ (quadrupole)
components of the power spectrum.
Confirmation of Eq.(1) would provide detailed information about the
kind of inflation that took place and, consequently, about the
evolution of the very early Universe.

For a given sum of $C_\ell^{(T)}$ and $C_\ell^{(S)}$ fixed by large
angular scale measurements, a larger tensor component reduces the
small-angle anisotropy.  Several other factors can have a
qualitatively similar effect (\eg increased tilt, $1-n_s$; decreased
baryon density, $\Omega_B$; and a nonstandard recombination
history), but there are {\it quantitative}
differences which we now describe.

To compute $C^{(T)}_{\ell}$, we evolve the distribution function,
${\bf f} ({\bf x}, {\bf q}, t) $, for photons at position ${\bf x}$ at
time $t$ with momentum ${\bf q}$, using first-order perturbation
theory of the general relativistic Boltzmann equation for radiative
transfer\cite{BE}, with a Thomson scattering source term.  Photon
polarization is included by making ${\bf f}$ a 4-dimensional vector
with components related to the Stokes parameters ($f_s$ with
$s=t,p,u,v$ correspond to the usual $I,Q,U,V$ Stokes notation) and
applying Chandrasekhar's development of the scattering source term for
Rayleigh (and thus Thomson) scattering in a plane parallel
atmosphere\cite{Cha}.
In the scalar case, only $f_t$ and
the `polarization' $f_p$ are needed, so 2 transfer equations are
required\cite{BE}. In the tensor case $f_u$ also does not vanish, but
it is related to $f_p$, so again only two perturbed transfer equations
turn out to be required. To describe these
equations, we introduce the relative perturbed distribution
functions\cite{BE} $\Delta_{s}^{(T)} = 4 \delta f_s/ (T_0
\; \partial {\bar f}/\partial T_0)$, where $T_0$ is the CMB
temperature and ${\bar f}$ is the unperturbed Planck distribution.

To evolve the coupled equations, both $\Delta_{s}^{(T)}$ and the
metric are expanded in plane waves. In the frame in which the
wavevector ${\bf k}$ is along the z-axis, the gravitational wave
degrees of freedom in the metric are the transverse traceless modes,
$h_{+}=h_{11}=-h_{22}$ and $h_{\times}=h_{12}=h_{21}$, which obey an
Einstein equation, the wave equation for free massless particles: $
{\ddot h}_\epsilon + 2{{\dot a}\over a}{\dot h}_\epsilon + k^2
h_\epsilon = 0 \, , \, \epsilon = +, \times $. Here the dot
denotes derivative {\it wrt} conformal time, $\tau = \int dt/a(t)$
where $a$ is the expansion factor, solved for by evolving the
Friedmann equation.

   The radiative transfer equations for the two gravity wave polarizations
separate, having an overall factor of $\cos(2\phi)$ for $\epsilon
=+$ and of $\sin(2\phi)$ for $\epsilon =\times$, where $(\theta , \phi)$
are the polar angles, which we remove by introducing new variables, following
Polnarev\cite{Pol}
\begin{equation}
\begin{array}{rcl}
\Delta_{t}^{(T)}&=&{\widetilde \Delta}_{t+}^{(T)}(1-\mu^2 )\cos 2\phi +
{\widetilde \Delta}_{t\times }^{(T)}(1-\mu^2 )\sin 2\phi
\\
\Delta_{p }^{(T)}& =&{\widetilde \Delta}_{p+ }^{(T)}
(1+\mu^2 )\cos 2\phi + {\widetilde \Delta}_{p\times }^{(T)}
(1+\mu^2 )\sin 2\phi\\
\Delta_{u }^{(T)}& =&- {\widetilde \Delta}_{u+  }^{(T)}
2\mu \sin 2\phi + {\widetilde \Delta}_{u\times  }^{(T)}
2\mu \cos 2\phi
\end{array}
\end{equation}
The combination ${\widetilde \Delta}_{p\epsilon }^{(T)}
+ {\widetilde \Delta}_{u \epsilon}^{(T)}$ is
unexcited by gravity waves, as is $\Delta_{v}^{(T)}$,
so the 4 Stokes radiative transfer equations reduce to two:
\begin{equation}
\begin{array}{rl}
\dot{{\widetilde \Delta}_{t\epsilon }^{(T)}} &= - i k \mu
{\widetilde \Delta}_{t\epsilon }^{(T)} - {\dot h}_{\epsilon}- a \sigma_T
n_e {\widetilde \Delta}_{t\epsilon}^{(T)} + a \sigma_T
n_e \Psi_\epsilon \ , \\
\dot{{\widetilde \Delta}_{p\epsilon}^{(T)}} &= - i k \mu {\widetilde
\Delta}_{p\epsilon}^{(T)}  - a \sigma_T
n_e {\widetilde \Delta}_{p\epsilon}^{(T)} - a \sigma_T
n_e \Psi_\epsilon \ , \\
\Psi_\epsilon &\equiv \Biggl\lbrack
{1\over10}{\widetilde \Delta}_{t\epsilon ,0}^{(T)}
-{1\over7}
{\widetilde \Delta}_{t\epsilon ,2}^{(T)}+ {3\over70}
{\widetilde \Delta}_{t\epsilon ,4}^{(T)}\\
& -{3\over 5}{\widetilde \Delta}_{p\epsilon ,0}^{(T)}
-{6\over 7}{\widetilde \Delta}_{p\epsilon ,2}^{(T)}
-{3\over 70}{\widetilde
\Delta}_{p\epsilon ,4}^{(T)} \Biggr\rbrack \; .
\end{array}
\end{equation}
Here $\sigma_T$ is the Thomson scattering cross-section, and $n_e$,
the free electron density,   is evolved
using a careful treatment of the recombination atomic
physics.

As in the scalar case\cite{BE}, we solve these equations
by expanding in Legendre polynomials, {\it e.g.},
${\widetilde \Delta}_{t\epsilon}=
\sum_{\ell} (2\ell +1){\widetilde \Delta}_{t\epsilon ,\ell }
P_{\ell} (\mu)$, converting (3) to a hierarchy of coupled equations.
Our solutions, ${\widetilde \Delta}_{t\epsilon ,\ell}(k,\tau ) $ and
${\widetilde \Delta}_{p\epsilon ,\ell}(k,\tau ) $,
can be combined into the power spectrum by summing over ${\bf k}$
and polarizations:
\begin{equation}
\begin{array}{rl}
C_\ell^{(T)} &={\pi\over 2} \sum_{{\bf k}}
(\ell -1)\ell (\ell +1) (\ell +2)
\Big\langle \vert {{\widetilde \Delta}_{tG,  \ell-2}\over
(2\ell -1)(2\ell +1)} \\
& -2 {{\widetilde \Delta}_{tG,  \ell} \over (2\ell -1)(2\ell +3)}
 +{{\widetilde \Delta}_{tG,  \ell+2} \over
(2\ell +1)(2\ell +3)} \vert^2 \Big\rangle \ ,
\end{array}
\end{equation}
where ${\widetilde \Delta}_{tG,\ell}\equiv ({\widetilde
\Delta}_{t+,\ell}- i{\widetilde \Delta}_{t\times ,\ell })/\sqrt{2}$.
A similar expression applies for the polarization power
spectrum.\cite{polar} A useful check is to  assume
recombination is sudden at
$\tau = \tau_r$. The free-streaming solution from $\tau_r$
to the present $\tau_0$ is then
${\widetilde \Delta}_{t\epsilon , \ell} =
(-i)^{\ell} \int_{\tau_r}^{\tau_0} d\tau j_\ell (k (\tau_0 -\tau ) ) {\dot
h}_\epsilon (\tau ) $. Substitution into eq.(4) gives the Abbott and
Wise\cite{AWSta} formula for $C_\ell^{(T)}$, with which our numerical
results
agree  for low $\ell$.

In this paper, we discuss results for standard cold dark matter (CDM)
models with normal recombination, although it is straightforward to
adapt the numerical codes to other cosmological models ({\it e.g.},
mixed hot and cold dark matter).  We let $n_s$, the ratio of
tensor-to-scalar quadrupole anisotropy, the baryon density $\Omega_B$
and the Hubble parameter\cite{hub} ${\rm h}$ freely vary.

Fig.~1 shows a CDM model
with $\Omega_B = 0.05$, ${\rm h}=0.5$ and
$n_s=0.85$, which accounts for the slight downward tilt in
$C_\ell^{(S)}$ at small $\ell$. (We plot $\ell (\ell +1) C_{\ell}$,
since it is flat for scale-invariant ($n_s=1$) $C_\ell^{(S)}$ at small
$\ell$.)  The sharp increase in $C_\ell^{(S)}$ for $\ell
\gta 50$ followed by increasingly damped oscillations
are due to  adiabatic compression of photons and Doppler
shifts during decoupling\cite{BE}.
The tensor mode behaves quite differently.  The
first
moments drop sharply; then the curve settles to a tilted spectrum
similar to the scalar case for  $5 \gta \ell
\gta 50$.  At $\ell \gta 50$, the tensor drops sharply just
as the scalar  rises.
 We have set
 $C_2^{(T)}/C_2^{(S)}\approx 1 $, the inflation prediction of
Eq.~1 for a tilt of $n_s=0.85$.
We also illustrate  how a scalar-only spectrum with
low $\Omega_B$ (\eg the dashed curve)
partially mimics the scalar-plus-tensor shape for $\Omega_B =0.05$,
(assuming a larger $C_2^{(S)}$). Clearly, precise
measurements are required to separately determine $C_2^{(T)}/C_2^{(S)}$
and $\Omega_B$.

\begin{figure}
\caption{
(a) Angular Power Spectra for the tilted standard CDM model shown for
tensor, scalar and the sum. The light dashed line is an
$\Omega_B=0.01$ model. (b) shows the filters for the experiments used
in this paper as examples (heavy lines).  The light lines are other
representative experiments.}
\end{figure}

The filter functions, which indicate the experimental sensitivity as a
function of $\ell$, are shown for various experiments in Fig.~1b. The
theoretical prediction for the {\it rms} fluctuations at each of the
points in an experiment is found by multiplying $C_\ell/4\pi$ by the
filter and summing over $\ell , m$. From Fig.1, we see that
large-angle experiments ({\it e.g.}, DMR\cite{dmr}, MIT\cite{mit} and
Tenerife\cite{ten}) are equally sensitive to tensor and scalar modes,
smaller angle experiments ({\it e.g.} SP89\cite{mein} and
OVRO\cite{ovro}) are sensitive mostly to scalar, while the
intermediate SP91\cite{gaier,schuster} can measure some tensor,
although predominantly scalar.

A quantitative experimental fit to cosmological parameters is obtained
by constructing likelihood functions ${\cal
L}_e(C_2^{(S)},C_2^{(T)},n_s,\Omega_B,{\rm h})$ for each experiment,
$e$, assuming Gaussian statistics.\cite{belm}
Assuming the experiments are statistically independent (because they
cover unrelated regions of the sky or very different angular
wavebands), we combine the ${\cal L}_e$'s to get the full likelihood,
${\cal L} = \prod_{e}{\cal L}_e$, as shown in Fig. 2.
For all but DMR, ${\cal L}_e$ is calculated using Bayesian
techniques\cite{belm} which take into account the removal of any
linear combinations of the data such as gradients or averages by
marginalizing over the coefficients, assuming uniform prior
probability distribution in these coefficients.  For SP91, the method
was extended to treat simultaneously the 4 frequency channels.  For
DMR, we used the Smoot \etal `90 A+B X 53 A+B' (quadrupole-subtracted)
correlation function\cite{dmr} with a Gaussian approximation for the
likelihood\cite{cth} A more complete analysis will only be possible
once the DMR data are released. In Fig.~2, we have taken
$\Omega_B=0.05(2{\rm h})^{-2}$, consistent with nucleosynthesis
limits\cite{bbn}, and ${\rm h}=0.5$.

Fig. 2(a) displays the DMR likelihood contours in the
$C_2^{(S)}$--$C_2^{(T)}$ plane for fixed $n_s$ ($0.85$), demonstrating
that DMR can measure $C_2^{(S)}+C_2^{(T)}$, but cannot discriminate
scalar and tensor.  A preferred tensor-scalar ratio does arise in
Fig.~2(b) as soon as we incorporate small angle data.  Figure 2(b)
combines the DMR data, the 4-frequency-channel data from a 9 point
strip\cite{gaier} and a 13 point strip\cite{schuster} in the SP91
experiment, as well as SP89\cite{mein} and OVRO\cite{ovro} data (which
give weak upper limits but no detections).  The SP91 scans appear to
have detections, but the signal may be contaminated by unknown
sources.

Fig.~2(b) is tantalizing but inconclusive evidence for a gravitational
wave contribution.  Future refinement can be anticipated by using
simulated data sets, constructed by taking single realizations of
theoretical signals and adding experimental noise associated with
statistical errors (but no systematic errors).  In Figure 2(c) and
2(d), the input signal is for a standard CDM model with
$n_s=0.85$ and equal $C_2^{(T)}$ and $C_2^{(S)}$ ($[7.5\times
10^{-6}]^2$). We then simulate a suite of plausible near-future
experiments: DMR with 4-year error bars; six 13 point strips from an
SP91 configuration ($18$-$27 \mu K$ error bars for each of the 4
frequency channels); six 9 point strips from an SP89\cite{mein}
configuration; and an OVRO22 configuration ($7^\prime$ beam,
$22^\prime$ double-difference throw, with $25 \mu K$ error bars).
Reduced ($\approx 15 \mu K$) error bars were taken for SP89 to
represent ongoing or planned experiments with beams $\sim 0.5^\circ$
which, with multifrequency observations, can achieve these
sensitivities\cite{max}.


  Fig.~2(d) shows two projections onto the $C_2^{(S)}$--$n_s$ plane.
The heavy contours are the likelihood if $C_2^{(T)}/C_2^{(S)}$ is
restricted to the trajectories predicted by inflation, Eq.~(1). The
maximum lies within 10\% of the input signal.  In contrast, the light
curves show the contours when the (unrestricted) maximum likelihood
value for the given $C_2^{(S)}$ and $n_s$ is taken. The extended 1
sigma band along $C_2^{(T)}/C_2^{(S)}\approx (2-5.2(1-n_s))^2$
indicates an inability to distinguish large tilt from large tensor
component for these sensitivities.  The band runs across the inflation
prediction (heavy lines), intersecting in a narrow range about $n_s
\approx 0.83$, very close to the input value.

We conclude that current and near-future anisotropy experiments are
unable by themselves to definitively test inflation (Eq.~1) or
determine the gravitational wave contribution to the CMB.  For the
short-term, conclusions can only be drawn by adding extra assumptions
and/or other data.  For example, we have already shown in Figs.~2(c,d)
that, if Eq.~1 is {\it assumed}, $n_s$ and the gravitational wave
imprint can be determined to within 2 sigma.  Alternatively, other
cosmological constraints can be invoked. For example, a variety of
arguments imply that the {\it rms} amplitude of the density
fluctuations on scales of $8 \hmpc$ ($\sigma_8$), which is used to
measure the amount of nonlinear dynamics in large-scale structure
calculations, cannot lie outside of the range 0.45 and 1 for CDM
models\cite{ebw,Abffo}; this translates to the shaded region
 in Fig.~2(d).  If one assumes a model of galaxy
clustering with linear biasing, a tilt $n_s \lta 0.65$ is required for
standard CDM models to reproduce the galaxy correlation
function\cite{Abffo} (which would exclude our input model). This
severe restriction may be relaxed with less simplistic CDM models of
galaxy formation, and in cosmologies with more power in the density
fluctuations than CDM has on large scales\cite{ebw}.

The long-term future is brighter. Extensive mapping of the microwave
sky on small and intermediate angular scales can lead to highly
accurate determinations of the spectrum; e.g., using the
filter functions of Fig.~1, we find that the limiting
cosmic variance uncertainty in $\Delta T/T$ is only a
few per cent for SP89 and SP91 configurations.
 Even at
large angles where the cosmic variance is higher, we find that 5\%
accuracy should be achievable. Hence, given
optimal experimental designs for
measuring large and small angle anisotropy, there should
be sufficient resolution for a
fully independent test for inflation, theories of large-scale
structure, and the imprint of gravitational waves.

\begin{figure}
\caption{Likelihood contour maps for scalar  \newline
($[C_2^{(S)}]^{1/2}/10^{-5}$) {\it v.s.} tensor
($[C_2^{(T)}]^{1/2}/10^{-5}$) amplitudes are shown in (a),(b) and (c)
for the standard CDM model with fixed ${n_s=0.85}$ tilt.  The light
curves are 1, 2 and 3 sigma lines, the heavy curve or `x' gives the
maximum likelihood. (a) DMR only.  (b) DMR plus the 9 and 13 point
SP91 data, along with SP89 plus OVRO. (c) shows the maps with
simulated large and small angle data consisting of DMR (with 4 year
error bars), six 13 point SP91 strips, six 9 point SP89 strips and one
OVRO22 strip. The mean CDM signal input into the simulated data is
denoted by the square.  (d) Shows 1,2 and 3 sigma likelihood contours
for the simulated data in $[C_2^{(S)}]^{1/2}$---$n_s$ space, constrained
to the
$C_2^{(T)}/C_2^{(S)}$ trajectory defined by Eq.(1) (solid) and the
unconstrained
maximum likelihood trajectory (dashed).  Shading indicates the
range for which
CDM models are not dynamically
viable.}
\end{figure}

We thank P. Lubin and M. Turner for useful conversations.  This
research was supported by the DOE at Penn (DOE-EY-76-C-02-3071), NSERC
at Toronto, the SERC at Oxford and the Canadian Institute for Advanced
Research.


\begin{references}
\bibitem{polar} We find the polarization induced by
the tensor mode is negligibly small ($\ll 1\%$),
whereas models with early reionization have
significant polarization, an added discriminant that
we will not explore here.
\bibitem{cth}
The Gaussian variance for the likelihood adds to the observational
errors\cite{dmr} an all-sky cosmic variance and a (small) signal-noise
interference variance computed assuming a uniform pixel noise which
best-fits the DMR error bars.
\end{references}
\end{document}